\documentclass[twocolumn,showpacs,preprintnumbers]{revtex4}
\usepackage{graphicx}
\usepackage{amssymb}

\def\thorn{I\kern-0.4em\raise0.35ex\hbox{\it o}}

\begin{document}
\title{Spatial and null infinity via advanced and retarded conformal factors}
\author{Sean A. Hayward}
\affiliation{Department of Science Education, Ewha Womans University,
Seodaemun-gu, Seoul 120-750, Korea\\
{\tt hayward@mm.ewha.ac.kr}}
\date{14th September 2003}

\begin{abstract}
A new approach to space-time asymptotics is presented, refining Penrose's idea
of conformal transformations with infinity represented by the conformal
boundary of space-time. Generalizing examples such as flat and Schwarzschild
space-times, it is proposed that the Penrose conformal factor be a product of
{\em advanced and retarded conformal factors}, which asymptotically relate
physical and conformal null (light-like) coordinates and vanish at future and
past null infinity respectively, with both vanishing at spatial infinity. A
correspondingly refined definition of asymptotic flatness at both spatial and
null infinity is given, including that the conformal boundary is locally a
light cone, with spatial infinity as the vertex. It is shown how to choose the
conformal factors so that this asymptotic light cone is locally a metric light
cone. The theory is implemented in the spin-coefficient (or null-tetrad)
formalism by a simple joint transformation of the spin-metric and spin-basis
(or metric and tetrad). The advanced and retarded conformal factors may be used
as expansion parameters near the respective null infinity, together with a
dependent expansion parameter for both spatial and null infinity, essentially
inverse radius. Asymptotic regularity conditions on the spin-coefficients are
proposed, based on the conformal boundary locally being a smoothly embedded
metric light cone. These conditions ensure that the Bondi-Sachs energy-flux
integrals of ingoing and outgoing gravitational radiation decay at spatial
infinity such that the total radiated energy is finite, and that the
Bondi-Sachs energy-momentum has a unique limit at spatial infinity, coinciding
with the uniquely rendered ADM energy-momentum.
\end{abstract}
\maketitle

\section{Introduction}
Space-time asymptotics, the study of isolated gravitational systems at large
distances, is one of the theoretical pillars of General Relativity, whereby one
can define physically important quantities which are difficult to capture in
general, such as gravitational radiation, the energy flux of gravitational
radiation and the active gravitational mass-energy of the system. The discovery
of the Bondi-Sachs energy-loss equation \cite{B,BBM,S}, relating the change in
total mass-energy to the energy flux of outgoing gravitational radiation, just
as for electromagnetic radiation, marked a transition from an epoch where some
argued that Einstein's original prediction of gravitational radiation rested on
mathematical artefacts, to the present epoch where there is significant
investment in expectations of detecting gravitational radiation from
astrophysical sources.

The early work on space-time asymptotics was brilliantly reformulated by
Penrose \cite{P,PR}, who realized how to handle infinite distances and times in
a mathematically finite way. The space-time metric is mapped to a multiple of
itself, the factor tending to zero in such a way that the new metric is finite
and well behaved at the physical space-time infinity. The mapping is described
as conformal since it preserves space-time angles and therefore causal
relations, while changing distances and durations. In the conformal picture,
infinity becomes a finite boundary where one can derive exact formulas,
encapsulating physical laws which were otherwise approximate or limiting.

Penrose's conformal theory rapidly became the standard framework for space-time
asymptotics at null (light-like) infinity, and was quite thoroughly implemented
using the spin-coefficient or null-tetrad formalism \cite{PR,NP,NU,GHP,NT}.
However, it does not cover spatial infinity, described initially by the ADM
method \cite{ADM} and by a spatial version of Penrose's method by Geroch
\cite{G}. The essential problem is to describe spatial and null infinity in a
unified way. Ashtekar and coworkers \cite{AH,AM,A} proposed and investigated
such a definition, realizing that spatial infinity is generally a directional
singularity of the conformal metric. However, there still seems to be no
practical calculational formalism. Some physically important issues are whether
the gravitational radiation decays near spatial infinity such that the
Bondi-Sachs energy exists and coincides with the ADM energy at spatial
infinity, and whether initial data on an asymptotically flat spatial
hypersurface (or past null infinity) determines final data at future null
infinity. For a more recent perspective, see e.g.\ Friedrich \cite{F}.

This article presents a quite simple, natural unification of spatial and null
infinity, based on a re-examination of Penrose's original insights. The key new
idea is that Penrose's conformal factor should be a product of advanced and
retarded conformal factors, which do the work of the conformal factor at future
and past null infinity respectively, while cooperating at spatial infinity so
that it is the vertex of a light cone, generating null infinity. These factors,
squared, differentially relate physical and conformal null coordinates near the
respective null infinity, up to the relevant coordinate freedom.

The basic idea is most easily apprehended in examples, described in \S II. \S
III gives a definition of asymptotic flatness intended to encapsulate the new
refinements, and shows that conformal infinity is locally a metric light cone,
thereby determining standard coordinates. \S IV shows how to implement the
theory in the spin-coefficient (or null-tetrad) formalism, using a simple joint
transformation of the spin-metric and spin-basis (or metric and null tetrad).
\S V identifies appropriate expansion parameters and proposes asymptotic
regularity conditions at both null and spatial infinity, based on the
asymptotic light cone being smoothly embedded. \S VI uses the Hawking
quasi-local energy \cite{H} to study the Bondi-Sachs energy flux (of ingoing
and outgoing gravitational radiation) and total energy at both null and spatial
infinity, and the ADM energy. \S VII concludes.

\section{Advanced and retarded conformal factors}
For flat space-time, the line element may be written as
\begin{equation}
ds^2=r^2dS^2+dr^2-dt^2
\end{equation}
where $dS^2$ is a line element for the unit sphere. In terms of dual-null (or
characteristic) coordinates
\begin{equation}
2\xi^\pm=t\pm r
\end{equation}
it becomes
\begin{equation}
ds^2=(\xi^+-\xi^-)^2dS^2-4d\xi^+d\xi^-.
\end{equation}
To approach infinity, the physical null coordinates $\xi^\pm$ are transformed
to conformal null coordinates $\psi^\pm$ by
\begin{equation}
\xi^\pm=\tan\psi^\pm.
\end{equation}
This describes sterographic projection, used to map the Earth on flat paper: if
one projects from the north pole of a unit sphere onto the equatorial plane,
with $\psi$ the angle from the vertical, then $\xi=\tan\psi$ is the distance
from the centre of a point in the plane. Thus the infinite plane is mapped into
the compact sphere, with the north pole representing infinity; the sphere
describes the whole plane, plus its infinity. The formula also occurs in
complex analysis as a way of using the Riemann sphere to represent the Argand
plane, plus a point at infinity.

The novel point concerns the derivatives occurring in the null coordinate
transformations,
\begin{equation}
(\omega^\pm)^2=\frac{d\psi^\pm}{d\xi^\pm}=\cos^2\psi^\pm
\end{equation}
which vanish at the relevant infinity. The line element written in $\psi^\pm$
coordinates therefore contains a term $4d\psi^+d\psi^-/(\omega^+\omega^-)^2$.
To obtain a metric which is regular at infinity, one can multiply by the
discrepancy, the square of
\begin{equation}\label{Omega}
\Omega=\omega^+\omega^-.
\end{equation}
Then
\begin{equation}
\Omega^2ds^2=\sin^2(\psi^+-\psi^-)dS^2-4d\psi^+d\psi^-.
\end{equation}
A final transformation
\begin{equation}\label{rhotau}
2\psi^\pm=\tau\mp\rho
\end{equation}
puts this in the form
\begin{equation}
\Omega^2ds^2=\sin^2\rho\,dS^2+d\rho^2-d\tau^2
\end{equation}
which is the standard line element of the Einstein static universe, a metric
$S^3\times R$. The physical space-time metric $g$ has been mapped to the
conformal metric $\Omega^2g$ and the physical space-time manifold has been
mapped to the conformal coordinate range $\psi^\pm\in(-\pi/2,\pi/2)$. Thus
physical infinity has been mapped to the boundary of this region. This
conformal boundary consists of two null hypersurfaces connected by three
points: future null infinity $\Im^+$, $\psi^+=\pi/2$,
$\psi^-\in(-\pi/2,\pi/2)$; past null infinity $\Im^-$, $\psi^-=-\pi/2$,
$\psi^+\in(-\pi/2,\pi/2)$; future temporal infinity $i^+$,
$\psi^+=\psi^-=\pi/2$; past temporal infinity $i^-$, $\psi^+=\psi^-=-\pi/2$;
and spatial infinity $i^0$, $\psi^+=-\psi^-=\pm\pi/2$. The two null
hypersurfaces locally have the structure of light cones, with the three points
each being vertices. This initially bizarre sixties spacewarp led Penrose
\cite{P,PR} to the definition of conformal infinity as essentially $\Omega=0$,
$\nabla\Omega\not=0$, the physical metric $g$ admitting a well behaved
conformal metric $\Omega^2g$, $\Omega$ being called the conformal factor.

Nevertheless, there is more structure here. Firstly, spatial infinity is the
vertex of a conformal light cone consisting of null infinity. Secondly,
$\Im^\pm$ are given by $\omega^\pm=0$, $\omega^\mp\not=0$, with
$\omega^+=\omega^-=0$ at spatial infinity. Clearly there is more information in
the sub-factors $\omega^\pm$ than in the Penrose factor $\Omega$ alone. This
structure will be used to refine Penrose's definition of asymptotic flatness.
Henceforth $\omega^+$ and $\omega^-$ will be called the {\em advanced and
retarded conformal factors} respectively.

As a second example, the Schwarzschild space-time with mass $M$ in standard
coordinates is
\begin{equation}
ds^2=r^2dS^2+(1-2M/r)^{-1}dr^2-(1-2M/r)dt^2.
\end{equation}
This can be written in dual-null coordinates
\begin{equation}
2\xi^\pm=t\pm r_*
\end{equation}
where
\begin{equation}
r_*=r+2M\ln(r-2M)
\end{equation}
rescales $r$ such that $dr/dr_*=1-2M/r$. Then
\begin{equation}
ds^2=r^2dS^2-(1-2M/r)4d\xi^+d\xi^-
\end{equation}
where $r$ is implicitly determined as a function of $r_*=\xi^+-\xi^-$. As
before, the physical null coordinates $\xi^\pm$ may be transformed to conformal
null coordinates $\psi^\pm$ by, for instance, simple inversion:
\begin{equation}
\xi^\pm=-1/\psi^\pm.
\end{equation}
This is valid only in a neighbourhood of spatial infinity, but is simple enough
to constitute a local canonical transformation. The derivatives occurring in
the null coordinate transformations are
\begin{equation}
(\omega^\pm)^2=\frac{d\psi^\pm}{d\xi^\pm}=(\psi^\pm)^2
\end{equation}
and one can fix $\omega^\pm=\mp\psi^\pm$ so that $\omega^\pm>0$ in the physical
space-time, $\xi^\pm$ and $\psi^\pm$ being future-pointing. As before, the
metric in $\psi^\pm$ coordinates can be regularized by multiplying by the
square of $\Omega=\omega^+\omega^-$. Then
\begin{equation}
\Omega^2ds^2=(\psi^+\psi^-r)^2dS^2-4(1-2M/r)d\psi^+d\psi^-.
\end{equation}
Transforming as before by (\ref{rhotau}) puts this in the form
\begin{equation}
\Omega^2ds^2=(\rho r/r_*)^2dS^2+(1-2M/r)(d\rho^2-d\tau^2).
\end{equation}
Since $r_*/r\to1$ as $r\to\infty$, the conformal metric becomes flat at
infinity, with $\rho$ playing the role of conformal radius:
\begin{equation}
\Omega^2ds^2\to\rho^2dS^2+d\rho^2-d\tau^2.
\end{equation}
Thus infinity consists of a locally metric light cone with spatial infinity as
the vertex.

\begin{figure}
\includegraphics[width=5cm,height=7cm,angle=0]{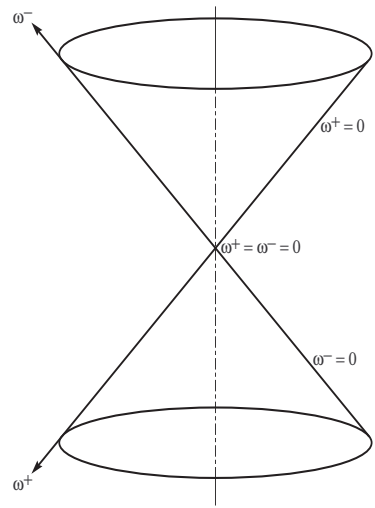}
 \caption{The light cone at infinity, depicting the advanced and retarded
 conformal factors $\omega^\pm$. The physical space-time ($\omega^+>0$,
 $\omega^->0$) lies outside this asymptotic light cone. Radiation propagates in
 from past null infinity $\Im^-$ ($\omega^-=0$) and out to future null infinity
 $\Im^+$ ($\omega^+=0$). Spatial infinity $i^0$ ($\omega^+=\omega^-=0$) is the
 vertex. For non-zero total mass, beware the singularity beyond infinity.}
 \label{fig}
\end{figure}

Space-time has been turned inside out: spatial infinity has become a point and
the physical space-time lies outside its light cone, as depicted in
Fig.~\ref{fig}. In more detail, the original region $\xi^+>0$, $\xi^-<0$ has
been mapped to $\psi^+<0$, $\psi^->0$, or $\omega^\pm>0$. The conformal
boundary again locally consists of $\Im^\pm$, where $\omega^\pm=0$,
$\omega^\mp\not=0$, and spatial infinity $i^0$, where $\omega^+=\omega^-=0$.
The case $M=0$ was not excluded, so this also provides a conformal
transformation of flat space-time such that infinity is locally a metric light
cone at spatial infinity. The key points are the identification of the advanced
and retarded conformal factors $\omega^\pm$ as differentially relating physical
and conformal null coordinates, and their behaviour at null and spatial
infinity.

\section{The light cone at infinity}
The above insights are now converted into a definition of asymptotic flatness
at both null and spatial infinity. In this section alone, $g$ will be used for
the conformal metric and $\tilde g$ for the physical metric. Quantities without
tildes, e.g.\ $\nabla$, similarly refer to the conformal metric.

In this article, a space-time $(\tilde M,\tilde g)$ will be said to be
asymptotically flat if the following conditions hold. (i) There exists a
space-time $(M,g)$ with boundary $\Im=\partial M$ and  functions $\omega^\pm>0$
on $\tilde M$ such that $M=\tilde M\cup\Im$ and $g=(\omega^+\omega^-)^2\tilde
g$ in $\tilde M$. (ii) $\Im$ is locally a light cone with respect to $g$, with
vertex denoted by $i^0$ and future and past open cones denoted by $\Im^+$ and
$\Im^-$ respectively. (iii) $\omega^\pm=0$, $\omega^\mp\not=0$,
$\nabla\omega^\pm\not=0$ on $\Im^\pm$, and
$g^{-1}(\nabla\omega^+,\nabla\omega^-)\not=0$ on $\Im$. (iv) ($g,\omega^\pm$)
are $C^k$ on $M-i^0$, with $k=3$ sufficing for the purposes of this article.
Differentiability at $i^0$ is deferred to \S V.

One may call $\Im=\Im^+\cup\Im^-\cup i^0$ (locally) the asymptotic light cone,
or more prosaically, the {\em light cone at infinity} \cite{AH}. The definition
can be easily seen to recover the basic conditions of Penrose's definition of
asymptotic simplicity \cite{P,PR}, namely $\Omega=0$ and $\nabla\Omega\not=0$
on $\Im^\pm$, with the Penrose factor $\Omega$ as before (\ref{Omega}). The
causal restriction on (plain or weak) asymptotic simplicity has been replaced
with the light-cone condition. Note by continuity that $\omega^+=\omega^-=0$ at
$i^0$. The differentiability conditions at spatial infinity are notoriously
awkward: if $g$ is $C^1$ then the total (ADM) energy vanishes, while if $g$ is
only $C^0$, the total energy is generally ill-defined \cite{AH,AM,A}. Readers
unable to proceed otherwise can provisionally assume the Ashtekar-Hansen
differentiability conditions at $i^0$. Essentially, a non-zero mass forces
radial derivatives of the conformal metric to be discontinuous at spatial
infinity, even for the Schwarzschild metric, while angular derivatives should
be much smoother. A geometrical way to understand this stems from the fact that
the positive-mass Schwarzschild space-time can be conformally extended through
$\Im^\pm$ to smoothly include a negative-mass Schwarzschild space-time whose
$(\Im^\mp,i^\mp)$ coincide with the physical $(\Im^\pm,i^0)$ \cite{PR}. Since
the negative-mass Schwarzschild space-time has a naked curvature singularity
extending from its $i^-$ to $i^+$, this {\em singularity beyond infinity}
(Fig.~\ref{fig}) piercing physical $i^0$ can be understood as forcing spatial
infinity to be a directional singularity of the conformal metric.

The analysis of Geroch \cite{G} for $\Im^\pm$, summarized together with the
Ashtekar-Hansen analysis by Wald \cite{W}, can be taken as a guide to the
following refinements. Firstly, the definition allows considerable gauge
freedom in the conformal factors $\omega^\pm$, namely
$\omega^\pm\mapsto\alpha_\pm\omega^\pm$ for any $C^k$ functions $\alpha_\pm>0$
on $M-i^0$. This gauge freedom allows the conformal metric of $\Im$ to be
locally fixed as that of a metric cone, as follows. The starting point is a
standard expression for the physical Ricci tensor $\tilde R$ in terms of the
conformal Ricci tensor $R$:
\begin{equation}\label{Ricci}
\Omega\tilde R=\Omega R+2\nabla\otimes\nabla\Omega
+(\nabla^2\Omega-3\Omega^{-1}g^{-1}(\nabla\Omega,\nabla\Omega))g.
\end{equation}
Applying either the vacuum Einstein equation $\tilde R=0$ or the full Einstein
equation with suitable fall-off of the matter fields, one sees that
$\Omega^{-1}g^{-1}(\nabla\Omega,\nabla\Omega)$ is smooth ($C^{k-2}$) at
$\Im^\pm$ and therefore that $\nabla\Omega$ is null there. Further, expanding
\begin{eqnarray}
&&\frac{g^{-1}(\nabla\Omega,\nabla\Omega)}{\Omega}
=2g^{-1}(\nabla\omega^+,\nabla\omega^-)+{}\\&&\qquad\qquad\quad
\frac{\omega^-}{\omega^+}g^{-1}(\nabla\omega^+,\nabla\omega^+)
+\frac{\omega^+}{\omega^-}g^{-1}(\nabla\omega^-,\nabla\omega^-)\nonumber
\end{eqnarray}
and using the above definition, one finds that
$(\omega^\mp/\omega^\pm)g^{-1}(\nabla\omega^\pm,\nabla\omega^\pm)$ are smooth
at $\Im^\pm$ and therefore that the 1-forms
\begin{equation}
n^\pm=\pm2\nabla\omega^\pm
\end{equation}
are null at $\Im^\pm$, where the signs are chosen so that the vectors
$g^{-1}(n^\pm)$ are future pointing, taking the choice $\omega^\pm>0$ in
$\tilde M$ and the $({-}{+}{+}{+})$ metric convention. The gauge freedom
$\omega^\pm\mapsto\alpha_\pm\omega^\pm$ can then be used to set
$g^{-1}(\nabla\omega^\pm,\nabla\omega^\pm)/\omega^\pm=0$ on $\Im^\pm$, by an
argument given by Geroch for $\Omega$. This still leaves $\omega^\mp$ free on
$\Im^\pm$. The normalization condition, which now reads
$g^{-1}(\nabla\omega^+,\nabla\omega^-)>0$ on $\Im$, expresses that the
hypersurfaces of constant $\omega^\pm$ intersect $\Im^\pm$ transversely; then
the gauge freedom allows one to choose them to intersect in a null direction,
$g^{-1}(\nabla\omega^\mp,\nabla\omega^\mp)=0$ on $\Im^\pm$. Thus we have two
independent null directions $g^{-1}(n^\pm)$ at $\Im^+$ and $\Im^-$. The local
null rescaling freedom of a null hypersurface can then be used to further
adjust $\omega^\mp$ on $\Im^\pm$ so that the null directions are relatively
normalized: $g^{-1}(\nabla\omega^+,\nabla\omega^-)=1/2$ on $\Im^\pm$. All the
above conditions extend to $i^0$ by continuity. Using $\approx$ to denote
equality on $\Im$ in a neighbourhood of $i^0$, the gauge conditions are
\begin{eqnarray}\label{gauge}
g^{-1}(\nabla\omega^\pm,\nabla\omega^\pm)/\omega^\pm&\approx&0\\
g^{-1}(\nabla\omega^+,\nabla\omega^-)&\approx&1/2.
\end{eqnarray}
Together they imply that
\begin{equation}
g^{-1}(\nabla\Omega,\nabla\Omega)/\Omega\approx1
\end{equation}
which shows that $\Omega$ is spatial in a neighbourhood of $\Im$. The gauge
condition usually chosen for $\Im^+$ is that this quantity vanishes.

Taking the trace and rearranging, the curvature relation (\ref{Ricci}) then
implies
\begin{equation}
2\nabla\otimes\nabla\Omega\approx g
\end{equation}
which again differs from the usual relation $\nabla\otimes\nabla\Omega=0$ on
$\Im^+$, which would imply that $\Im^+$ be a metric cylinder. Instead it
implies that $\Im$ is a metric cone, as follows. First note that there are now
preferred spatial sections of $\Im^\pm$, given by constant $\omega^\mp$.
Expanding the above relation,
\begin{equation}
g\approx2\omega^+\nabla\otimes\nabla\omega^-
+2\omega^-\nabla\otimes\nabla\omega^++4\nabla\omega^+\otimes\nabla\omega^-.
\end{equation}
The last term gives the part of the metric normal to the spatial sections of
$\Im$, while the first two terms are related to the expansions $\theta_\mp$ and
shears $\sigma_\mp$ of $\Im^\pm$ by $\bot2\nabla\otimes n^\pm=\theta_\mp
h+\sigma_\mp$ on $\Im^\pm$, where $h$ is the metric of the spatial sections and
$\bot$ denotes projection by $h$. Since
\begin{equation}
g\approx h+4\nabla\omega^+\otimes\nabla\omega^-
\end{equation}
one reads off $\theta_\mp=\pm2/\omega^\mp$, $\sigma_\mp=0$ on $\Im^\pm$.
However, this describes a metric light cone, as follows. The flat metric
\begin{equation}
ds^2=\rho^2dS^2+d\rho^2-d\tau^2
\end{equation}
in dual-null coordinates (\ref{rhotau}) is
\begin{equation}
ds^2=(\psi^--\psi^+)^2dS^2-4d\psi^+d\psi^-
\end{equation}
where the signs of $\psi^\pm$ have been chosen so that they are
future-pointing. Then $\sigma_\mp=0$ and
$\theta_\mp=2\partial_\mp\rho/\rho=2/\psi^\mp$ on $\Im^\pm$. Since the
expansions and shears, including vertex conditions, characterize the intrinsic
geometry of the light cone, this identifies the conformal metric with the flat
metric at $\Im$ if
\begin{equation}
\psi^\pm\approx\mp\omega^\pm.
\end{equation}
In general, $\psi^\pm$ can be taken as null coordinates with this gauge choice
at $\Im$. In summary, the conformal metric at $\Im$ has been locally fixed as
that of a metric light cone.

\section{Formalisms}
The most developed implementation of conformal infinity involves the
spin-coefficient formalism, also due to Penrose and coworkers \cite{NP,GHP},
which provides the most elegant formulation of various issues such as the Sachs
``peeling'' of the gravitational field near $\Im^+$ \cite{S,P,PR}. Henceforth
this formalism is employed, though an attempt is made to render the description
accessible to the uninitiated. Alternatively, the null-tetrad formalism
suffices for most purposes.

Henceforth equation numbers in triples will indicate equations in Penrose \&
Rindler \cite{PR}. It is convenient to modify some notation and summarize, as
follows. The basic geometrical object is a 2-form $\varepsilon$ (2.5.2) acting
on complex 2-spinors (spin-vectors), here called the {\em spin-metric}. It can
be regarded as a square root of the metric, which is expressed as a direct
product $g=\varepsilon\circ\bar\varepsilon$, where the bar denotes the complex
conjugate and the exact meaning of the product is given in abstract index
notation (3.1.9). An ordered pair of non-parallel spin-vectors $(o,\iota)$ is
here called a {\em spin-basis}. Expressing vectors as spin-vector dyads, this
defines a null tetrad, a vector basis (3.1.14):
\begin{equation}
(l,m,\bar m,l')=(o\bar o,o\bar\iota,\iota\bar o,\iota\bar\iota).
\end{equation}
Then $(l,l')$ are real null vectors, whereas $m$ is a complex vector encoding
transverse spatial vectors. In terms of the complex normalization factor
(2.5.46)
\begin{equation}
\chi=\varepsilon(o,\iota)
\end{equation}
one finds
\begin{equation}
g(l,l')=-g(m,\bar m)=\chi\bar\chi
\end{equation}
with the other eight independent (symmetric) contractions of the tetrad vectors
vanishing. Here the metric convention has switched to $({+}{-}{-}{-})$, as is
regrettably standard in the spin-coefficient formalism. Some conventional
factors of $\sqrt2$ are also entailed compared to the previous sections, e.g.\
physical null coordinates $(\xi,\xi')=\sqrt2(\xi^+,\xi^-)$ and conformal null
coordinates $(\psi,\psi')=\sqrt2(\psi^+,\psi^-)$.

The inverse metric can be written
\begin{equation}
g^{-1}=\frac2{\chi\bar\chi}(l\otimes l'-m\otimes\bar m)
\end{equation}
where $\otimes$ denotes the symmetric tensor product. The dual basis of 1-forms
(4.13.32) will be denoted by
\begin{equation}
(n,w,\bar w,n')=g(l',-\bar m,-m,l)
\end{equation}
so that
\begin{equation}
n(l)=w(m)=\bar w(\bar m)=n'(l')=1
\end{equation}
with the other twelve such contractions vanishing. Then the metric can be
written
\begin{equation}
g=2\chi\bar\chi(n\otimes n'-w\otimes\bar w).
\end{equation}
The tetrad covariant derivative operators (4.5.23) are denoted by
\begin{equation}
(D,\delta,\delta',D')=(\nabla_l,\nabla_m,\nabla_{\bar m},\nabla_{l'}).
\end{equation}
The complex spin-coefficients
$(\kappa,\sigma,\rho,\tau,\varepsilon,\beta,\alpha,\gamma)$ and
$(\kappa',\sigma',\rho',\tau',\varepsilon',\beta',\alpha',\gamma')$ encode the
Ricci rotation coefficients and can be defined by (4.5.26--27) as
\begin{eqnarray}
D(o,\iota)&=&(\varepsilon o-\kappa\iota,\gamma'\iota-\tau'o)\\
\delta(o,\iota)&=&(\beta o-\sigma\iota,\alpha'\iota-\rho'o)\\
\delta'(o,\iota)&=&(\alpha o-\rho\iota,\beta'\iota-\sigma'o)\\
D'(o,\iota)&=&(\gamma o-\tau\iota,\varepsilon'\iota-\kappa'o).
\end{eqnarray}
Other notation will be taken as in Penrose \& Rindler \cite{PR}.

It is convenient to choose null coordinates $(\xi,\xi')$ such that
$(n,n')=(d\xi,d\xi')$, which implies the dual-null gauge conditions \cite{bhs}
\begin{eqnarray}\label{dualnull0}
&&0=\kappa=\rho-\bar\rho=\varepsilon+\bar\varepsilon=\tau-\bar\alpha-\beta\\
\label{dualnull1} &&0=\kappa'=\rho'-\bar\rho'=\varepsilon'+\bar\varepsilon'
=\tau'-\bar\alpha'-\beta'.
\end{eqnarray}
It is also convenient to use Penrose's complex stereographic coordinate
(1.2.10)
\begin{equation}
\zeta=e^{i\phi}\cot(\theta/2)
\end{equation}
where $(\theta,\phi)$ are standard spherical polar coordinates. Then $(w,\bar
w)=(d\bar\zeta/P,d\zeta/\bar P)$, (4.14.29), where (4.14.31)
\begin{equation}
P=\delta\bar\zeta.
\end{equation}
In summary, we have preferred coordinates $(\xi,\zeta,\bar\zeta,\xi')$ such
that the 1-form basis is
\begin{equation}
(n,w,\bar w,n')=(d\xi,d\bar\zeta/P,d\zeta/\bar P,d\xi').
\end{equation}
The above quantities will refer to the physical metric $g$ and corresponding
quantities for the conformal metric $\hat g$ will be denoted by hats.

The physical coordinates are now to be transformed to conformal coordinates
$(\psi,\zeta,\bar\zeta,\psi')$, such that the angular coordinates
$\hat\zeta=\zeta$ are preserved, while the null coordinates transform to
$(\hat\xi,\hat\xi')=(\psi(\xi),\psi'(\xi'))$, with derivatives
\begin{equation}
(\omega^2,\omega'^2)=\left(\frac{d\psi}{d\xi},\frac{d\psi'}{d\xi'}\right).
\end{equation}
Taking $\xi$ as an advanced (outgoing) null coordinate and $\xi'$ as a retarded
(ingoing) null coordinate, this means that $\omega$ and $\omega'$ are the
advanced and retarded conformal factors respectively, assumed positive in the
space-time. For the canonical example of inversion,
$(\psi,\psi')=(-2/\xi,-2/\xi')$ with the current convention, yielding
$(\omega,\omega')=(-\psi,\psi')/\sqrt2$. The Penrose conformal factor is
\begin{equation}
\Omega=\omega\omega'
\end{equation}
and the physical spin-metric $\varepsilon$ is correspondingly transformed to
the conformal spin-metric
\begin{equation}\label{spin}
\hat\varepsilon=\omega\omega'\varepsilon
\end{equation}
and the physical metric $g$ to the conformal metric $\hat
g=\hat\varepsilon\circ\hat{\bar\varepsilon}$:
\begin{equation}\label{metric}
\hat g=(\omega\omega')^2g.
\end{equation}
By comparing the explicit forms
\begin{eqnarray}\label{g}
g&=&2|\chi|^2\left(d\xi\otimes d\xi'-\frac{d\zeta\otimes
d\bar\zeta}{|P|^2}\right)\\
\hat g&=&2|\hat\chi|^2\left(d\psi\otimes d\psi'-\frac{d\zeta\otimes
d\bar\zeta}{|\hat P|^2}\right)
\end{eqnarray}
one reads off
\begin{eqnarray}\label{chi}
\hat\chi&=&\chi\\\label{P}\hat P&=&P/\omega\omega'.
\end{eqnarray}
Then the conformal 1-form basis
\begin{equation}
(\hat n,\hat w,\hat{\bar w},\hat n')=(d\psi,d\bar\zeta/\hat P,d\zeta/\hat{\bar
P},d\psi')
\end{equation}
is related to the physical 1-form basis by
\begin{equation}
(\hat n,\hat w,\hat{\bar w},\hat
n')=(\omega^2n,\omega\omega'w,\omega\omega'\bar w,\omega'^2n').
\end{equation}
Since
\begin{equation}
\hat g^{-1}=g^{-1}/(\omega\omega')^2
\end{equation}
the physical null tetrad
\begin{equation}
(l,m,\bar m,l')=g^{-1}(n',-\bar w,-w,n)
\end{equation}
and the conformal null tetrad
\begin{equation}
(\hat l,\hat m,\hat{\bar m},\hat l')=\hat g^{-1}(\hat n',-\hat{\bar w},-\hat
w,\hat n)
\end{equation}
are then related by
\begin{equation}\label{tetrad}
(\hat l,\hat m,\hat{\bar m},\hat
l')=\left(\frac{l}{\omega^2},\frac{m}{\omega\omega'},\frac{\bar
m}{\omega\omega'},\frac{l'}{\omega'^2}\right).
\end{equation}
Remarkably, this is just the behaviour of the null tetrad under a
transformation of spin-basis $(o,\iota)$ to
\begin{equation}\label{frame}
(\hat o,\hat\iota)=(o/\omega,\iota/\omega').
\end{equation}
Such transformations $(o,\iota)\mapsto(\lambda o,\lambda'\iota)$, with
$(\lambda,\lambda')$ generally complex, (4.12.2), are well understood and led
to the concept of weighted scalars (4.12.9) and the compacted spin-coefficient
formalism \cite{GHP}. In summary, it has been shown that {\em the desired
conformal transformation is given by a simultaneous transformation of the
spin-metric and spin-basis}, (\ref{spin}) and (\ref{frame}). This can be so
without the conformal factors $(\omega,\omega')$ exactly relating physical and
conformal null coordinates as above, due to the considerable freedom in the
conformal factors.

The physical and conformal tetrad derivative operators are related by
\begin{equation}\label{deriv}
(\hat D,\hat\delta,\hat\delta',\hat D')=
\left(\frac{D}{\omega^2},\frac{\delta}{\omega\omega'},
\frac{\delta'}{\omega\omega'},\frac{D'}{\omega'^2}\right).
\end{equation}
The relations between the weighted spin-coefficients are obtained
straightforwardly from (5.6.15) as
\begin{eqnarray}\label{sc0}
\hat\kappa&=&\kappa\omega'/\omega^3\\
\hat\sigma&=&\sigma/\omega^2\\
\hat\rho&=&(\rho-D\log\omega\omega')/\omega^2\\
\hat\tau&=&(\tau-\delta\log\omega\omega')/\omega\omega'\\
\hat\tau'&=&(\tau'-\delta'\log\omega\omega')/\omega\omega'\\
\hat\rho'&=&(\rho'-D'\log\omega\omega')/\omega'^2\\
\hat\sigma'&=&\sigma'/\omega'^2\\\label{sc1}
\hat\kappa'&=&\kappa'\omega/\omega'^3.
\end{eqnarray}
They are much simpler than the expressions (5.6.25) or (5.6.27) used previously
to describe conformal transformations, due to the separation of the Penrose
conformal factor into advanced and retarded conformal factors. The components
(4.11.6)
\begin{eqnarray}
\Psi_0&=&\Psi(o,o,o,o)/|\chi|^2\\
\Psi_1&=&\Psi(o,o,o,\iota)/|\chi|^2\\
\Psi_2&=&\Psi(o,o,\iota,\iota)/|\chi|^2\\
\Psi_3&=&\Psi(o,\iota,\iota,\iota)/|\chi|^2\\
\Psi_4&=&\Psi(\iota,\iota,\iota,\iota)/|\chi|^2
\end{eqnarray}
of the conformally invariant Weyl curvature spinor $\Psi=\hat\Psi$ (4.6.35)
transform even more simply as
\begin{eqnarray}\label{weyl0}
\hat\Psi_0&=&\Psi_0/\omega^4\\
\hat\Psi_1&=&\Psi_1/\omega^3\omega'\\
\hat\Psi_2&=&\Psi_2/\omega^2\omega'^2\\
\hat\Psi_3&=&\Psi_3/\omega\omega'^3\\\label{weyl1}
\hat\Psi_4&=&\Psi_4/\omega'^4.
\end{eqnarray}
One can also generalize the concept of {\em conformal density} (5.6.32) to a
quantity $\eta$ which transforms as
\begin{equation}
\hat\eta=\eta/\omega^k\omega'^{k'}
\end{equation}
where $\{k,k'\}$ is the {\em conformal weight}, with signs chosen for
convenience. Then the shears $(\sigma,\sigma')$ have conformal weights
$\{2,0\}$ and $\{0,2\}$ respectively. Also $\tau-\bar\tau'$ has conformal
weight $\{1,1\}$. Generally $\tau+\bar\tau'$ is not a conformal density, but it
is so in the dual-null gauge (\ref{dualnull0}--\ref{dualnull1}), which yields
\cite{bhs}
\begin{equation}\label{dchi}
\tau+\bar\tau'=\delta\log\chi\bar\chi.
\end{equation}
Then $(\tau,\tau')$ both have conformal weight $\{1,1\}$.

It should also be remarked that this implementation can be achieved purely in
the null-tetrad formalism, without ever mentioning spinors. For instance, the
weighted spin-coefficients can be expressed as (4.5.22)
\begin{eqnarray}
(\kappa,\sigma,\rho,\tau)&=&(Dn',\delta n',\delta'n',D'n')(m)/|\chi|^2\\
(\tau',\rho',\sigma',\kappa')&=&(Dn,\delta n,\delta'n,D'n)(\bar m)/|\chi|^2.
\end{eqnarray}
The components of the Weyl and Ricci spinors can also be expressed in terms of
the null tetrad (4.11.9--10), as can the weighted derivative operators
$(\thorn,\eth,\eth',\thorn')$ (4.12.15) acting on tensors, and consequently all
of the compacted spin-coefficient equations (4.12.32). The only spinorial
remnant is the phase of $\chi$, which does not enter tensorial expressions. In
this purely tensorial view, {\em the desired conformal transformation is given
by a simultaneous transformation of the metric and null tetrad}, (\ref{metric})
and (\ref{tetrad}).

\section{Asymptotic regularity}
It is now possible to investigate the asymptotic behaviour of the space-time
near both spatial and null infinity. While a detailed set of asymptotic
expansions is not derived in the present article, a leading-order analysis
suffices to derive some key results.

Fixing the null coordinates $(\psi^+,\psi^-)=(\psi,\psi')/\sqrt2$ on $\Im$ as
in \S III so that it is locally a metric light cone, the conformal metric
spheres are propagated into $M$ by the vectors $(\hat l,\hat l')$, forming a
preferred two-parameter family of transverse spatial surfaces, close to metric
spheres, in a neighbourhood of $\Im$. It is also convenient to fix the
conformal freedom by
\begin{equation}
(\omega,\omega')=(-\psi,\psi')/\sqrt2
\end{equation}
so that the advanced and retarded conformal factors themselves can be used as
conformal null coordinates, and as expansion parameters near $(\Im^+,\Im^-)$
respectively. For expansions near both spatial and null infinity, a useful
combination is
\begin{equation}
u=\frac{\omega\omega'}{\omega+\omega'}.
\end{equation}
This parameter has the properties of being linear in spatial linear
combinations of $(\omega,\omega')$ near $i^0$, with $u\sim\omega$ at $\Im^+$
and $u\sim\omega'$ at $\Im^-$, and that the constant-$u$ hypersurfaces are
hyperboloids wrapping the asymptotic light cone. The expansion parameter $u$ is
also asymptotically the inverse radius $1/r$ of the transverse surfaces, as
follows. This in turn implies $\omega\sim1/r$ at $\Im^+$ and $\omega'\sim1/r$
at $\Im^-$.

Inspecting the form of the metric (\ref{g}), a radius function can be defined
by
\begin{equation}
r=|\chi||P_0|/|P|
\end{equation}
where (4.15.116)
\begin{equation}
P_0=(\zeta\bar\zeta+1)/\sqrt2
\end{equation}
refers to the unit sphere. The conformal radius
\begin{equation}
\hat r=|\hat\chi||P_0|/|\hat P|
\end{equation}
is then related by (\ref{chi})--(\ref{P}) as
\begin{equation}
\hat r=\omega\omega'r.
\end{equation}
However, for the metric spheres at $\Im$ this is just $\rho$ of \S II--III,
\begin{equation}
\hat r\approx\omega+\omega'
\end{equation}
and so one finds
\begin{equation}
u\sim1/r.
\end{equation}
Here and henceforth, $a\sim b$ means $a/b\approx1$, i.e.\ they have the same
leading-order asymptotic behaviour at $\Im$. When considering spatial or null
infinity separately, one can always use $u$ as the sole expansion parameter,
remembering that $(\omega,\omega')$ count as $(u,O(1))$ at $\Im^+$, $(O(1),u)$
at $\Im^-$ and $(O(u),O(u))$ at $i^0$ from spatial directions. When treating
the whole of $\Im$, it is convenient to use all three expansion parameters
explicitly.

Now one wishes to develop asymptotic expansions valid near both spatial and
null infinity. In the usual treatment of null infinity, one can derive the
asymptotic expansions from the definition of asymptotic simplicity \cite{PR}.
In the current context, this is not so, since the differentiability at $i^0$
was deliberately left open. This issue will now be examined in terms of the
spin-coefficients.

Using (\ref{deriv}), it is straightforward to calculate the useful expressions
\begin{eqnarray}
\hat D(\omega,\omega',u)&=&(-1,0,-u^2/\omega^2)/\sqrt2\\
\hat D'(\omega,\omega',u)&=&(0,1,u^2/\omega'^2)/\sqrt2\\
D(\omega,\omega',u)&=&(-\omega^2,0,-u^2)/\sqrt2\\
D'(\omega,\omega',u)&=&(0,\omega'^2,u^2)/\sqrt2.
\end{eqnarray}
With the asymptotic gauge choice
\begin{equation}
\chi\approx1
\end{equation}
the conformal convergences are found as
\begin{eqnarray}\label{rho0}
\hat\rho&=&-\frac{\hat D\hat r}{\hat r}\sim1/\sqrt2(\omega+\omega')\\
\label{rho1} \hat\rho'&=&-\frac{\hat D'\hat r}{\hat
r}\sim-1/\sqrt2(\omega+\omega').
\end{eqnarray}
Putting these results together and using the spin-coefficient transformations
(\ref{sc0})--(\ref{sc1}), the physical convergences are straightforwardly
calculated to have the leading-order behaviour
\begin{eqnarray}
\rho&\sim&-u/\sqrt2\\
\rho'&\sim&u/\sqrt2
\end{eqnarray}
which agree with standard expressions at $\Im^+$ \cite{PR,NU,NT}, but with a
symmetric relative normalization of the null tetrad.

For the remaining weighted spin-coefficients, which are all conformal densities
in the dual-null gauge (\ref{dualnull0}--\ref{dualnull1}), one can apply the
generalized peeling theorem (9.7.4). In the current context, this shows that a
scalar conformal density $\eta$ of weight $\{k,k'\}$ satisfies
$\eta=O(\omega^k)$ at $\Im^+$ and $\eta=O(\omega'^{k'})$ at $\Im^-$. This
constrains the possible asymptotic behaviour at $\Im^\pm$ but does not
determine it, nor the behaviour at $i^0$. It is tempting to conjecture the
minimal asymptotic behaviour $\eta=O(\omega^k\omega'^{k'})$ for the whole of
$\Im$, which turns out to be consistent with the following.

In any case, asymptotic behaviour of the conformal spin-coefficients is more
precisely determined by geometrical arguments, essentially concerning the
intrinsic and extrinsic curvature of a smoothly embedded metric light cone.
This has already been done above for the conformal convergences
(\ref{rho0}--\ref{rho1}). For the conformal shears, which vanish at the
respective null infinity, as shown in \S III or as (9.6.28), this suggests
$\hat\sigma'=O(\omega)$ at $\Im^+$. On the other hand, there is no reason for
$\hat\sigma'$ to vanish at $\Im^-$ or $i^0$, though it should be finite. With
corresponding conditions for $\hat\sigma$, assuming dependence on integral
powers of $(\omega, \omega',u)$ uniquely yields
\begin{eqnarray}
\hat\sigma&=&O(u/\omega)\\
\hat\sigma'&=&O(u/\omega').
\end{eqnarray}
To be more precise, a function $f(\omega,\omega',\theta,\phi)$ is said to be
{\em regular at} $\Im$ if its limits at $\Im^+$ ($\omega\to0$) and $\Im^-$
($\omega'\to0$) exist, together with the limits
\begin{equation}\label{regular}
f_0(\theta,\phi)=\lim_{\omega\to0}f(\omega,0,\theta,\phi)
=\lim_{\omega'\to0}f(0,\omega',\theta,\phi).
\end{equation}
To economize on notation, $a=O(b)$ applied at $\Im$ is here and henceforth
taken to mean that $a/b$ tends to a function regular at $\Im$. This allows the
limits at $i^0$ to depend on the angular direction, as is necessary e.g.\ for
$\Psi_2/u^3$ in the Kerr solution \cite{AH}.

For $(\hat\tau,\hat\tau')$, one can use the geometrical interpretation of
$\hat\tau-\hat{\bar\tau}'$ as the conformal twist of the dual-null foliation,
measuring the lack of commutativity of the null normal vectors $(l,l')$
\cite{mon}. This suggests that it should be $O(\omega\omega')$. Similarly
$\hat\tau+\hat{\bar\tau}'$ (\ref{dchi}) measures transverse derivatives of the
relative normalization of the null normals, suggesting similar behaviour. Then
\begin{eqnarray}
\hat\tau&=&O(\omega\omega')\\
\hat\tau'&=&O(\omega\omega').
\end{eqnarray}
In terms of the physical spin-coefficients, this yields
\begin{eqnarray}\label{shear0}
\sigma&=&O(u\omega)\\\label{shear1}
\sigma'&=&O(u\omega')\\\label{twist0}
\tau&=&O((\omega\omega')^2)\\\label{twist1}
\tau'&=&O((\omega\omega')^2).
\end{eqnarray}
The practical test of these geometrically motivated conditions is whether they
describe a class of space-times with desired physical properties, as considered
in the next section.

Summarizing this section: leading-order behaviour of the conformal weighted
spin-coefficients has been proposed according to the geometrical nature of
$\Im$ as a smoothly embedded metric light cone with respect to the conformal
metric. This directly determines the leading-order behaviour of the physical
weighted spin-coefficients
$(\kappa,\sigma,\rho,\tau,\tau',\rho',\sigma',\kappa')$. One might translate
the conditions back to differentiability conditions on the metric, which would
be stricter than previous suggestions \cite{AH,AM,A}, but such metric-level
requirements are not particularly illuminating.

Since the asymptotic regularity conditions necessarily allow angular dependence
at spatial infinity, it may sometimes be useful to expand $i^0$ from a point to
a sphere, defined by $\omega\to0$, $\omega'\to0$ and parametrized by
$(\theta,\phi)$. This is in contrast to previous descriptions of $i^0$ as a
hyperboloid \cite{G,AH,A}, depending also on the boost direction, or as a
cylinder \cite{F}. Here there is no detailed dependence on the boost direction
$\omega/\omega'$; to use Geroch's terminology of universal versus physical
structure \cite{G}, the boost dependence at $i^0$ is treated universally here,
described using $(\omega,\omega')$ in the given gauge, while the physical
dependence at $i^0$ is purely angular, encoded in the next section in functions
$(\chi_1,\rho_1,\rho'_1,\sigma_1,\sigma'_1)$ which are regular at $\Im$,
defined as above (\ref{regular}) to allow only angular dependence at $i^0$.
This seems to reflect the intuitive nature of spatial infinity as a large
time-independent sphere.

\section{Energy}
It is widely agreed that the most physically important discovery in space-time
asymptotics is the Bondi-Sachs energy-loss equation \cite{B,BBM,S}, whereby the
total mass-energy $E_{BS}$ decreases at $\Im^+$ according to \cite{PR,NT,mon}
$\sqrt2D'E_{BS}=-\oint{\tilde*}|N'|^2/4\pi$, where the integral is over a
sphere at $\Im^+$, ${\tilde*}1$ denotes the area form of a unit sphere and the
complex function $N'$ corresponds to what Bondi dubbed the news. In more
physical terminology, $|N'|^2$ is the conformal energy flux of the
gravitational radiation, meaning that the physical energy flux is $|N'|^2/r^2$
near $\Im^+$, the above integral giving the integrated energy flux through a
sphere near $\Im^+$. Now the news is $N'=\sigma'/u$ \cite{mon}, as will be
verified below. The asymptotic regularity condition (\ref{shear1}) then implies
$N'=O(\omega')$, so that it is not only finite at $\Im^+$, but decays near
$i^0$. Transforming the Bondi-Sachs energy-loss equation to use the conformal
null derivative (\ref{deriv}) yields $\hat D'E_{BS}=D'E_{BS}/\omega'^2=O(1)$.
Thus the gravitational radiation decays near spatial infinity in such a way
that the total energy flux, integrated over time, is finite. Moreover, the
prescribed fall-off of $\sigma'$ is the weakest which would achieve this
result. This property has been presented first because it follows so simply
from the regularity conditions, thereby verifying that the asymptotic behaviour
of the shears, proposed on geometrical grounds, is exactly right on physical
grounds.

It is also independent even of the existence of the total energy, for which one
needs to expand some spin-coefficients further, specifically
\begin{eqnarray}\label{exp0}
\chi-1&\sim&\chi_1u\\
-\sqrt2\rho/u-1&\sim&\rho_1u\\\label{exp1}
\sqrt2\rho'/u-1&\sim&\rho'_1u
\end{eqnarray}
where $(\chi_1,\rho_1,\rho'_1)$ are regular at $\Im$. Here the proposal is that
$u$ is the natural expansion parameter near both spatial and null infinity,
even though leading-order behaviour may depend on $(\omega,\omega')$
independently

To study energy near $\Im$, a useful quantity is the Hawking quasi-local
mass-energy \cite{H}, which can be written as \cite{mon}
\begin{equation}
E=\frac{A^{1/2}}{(4\pi)^{3/2}}\oint{*}\frac{K+\rho\rho'}{\chi\bar\chi}
\end{equation}
where the integral is over a transverse surface, ${*}1$ denotes the area form
of a transverse surface,
\begin{equation}
A=\oint{*}1
\end{equation}
is the area of a transverse surface and (4.14.20)
\begin{equation}\label{K}
K=\sigma\sigma'-\rho\rho'-\Psi_2+\Phi_{11}+\Pi
\end{equation}
is the complex curvature, satisfying a complex generalization of the
Gauss-Bonnet theorem (4.14.42--43):
\begin{equation}
\oint{*}\frac{K}{\chi\bar\chi}=\oint{*}\frac{\bar
K}{\chi\bar\chi}=2\pi(1-\gamma)
\end{equation}
where $\gamma$ is the genus of the transverse surface, vanishing in this
context. For the Schwarzschild space-time considered in \S II, one finds
$\chi=(1-2M/r)^{1/2}$ (phase irrelevant), $\rho=-Dr/r=-(1-2M/r)/\sqrt2r$,
$\rho'=-D'r/r=(1-2M/r)/\sqrt2r$, $K=1/2r^2$ and therefore $E=M$.

The Bondi-Sachs energy \cite{B,BBM,S} at $\Im^+$ can be expressed as
\cite{P,PR,NT,mon}
\begin{equation}
E_{BS}=\lim_{\omega\to0}\frac{A^{1/2}}{(4\pi)^{3/2}}
\oint{*}\frac{\sigma\sigma'-\Psi_2}{\chi\bar\chi}.
\end{equation}
In vacuum or with suitable fall-off of the matter terms $\Phi_{11}$ and $\Pi$,
(\ref{K}) shows that it can be written as the limit of the Hawking energy:
\begin{equation}\label{BS}
E_{BS}=\lim_{u\to0}E.
\end{equation}
Since
\begin{eqnarray}
{*}1&\sim&{\tilde*}1/u^2\\
A&\sim&4\pi/u^2
\end{eqnarray}
it is straightforward to see from the expansions (\ref{exp0}--\ref{exp1}) that
it is finite, given specifically by
\begin{equation}
E_{BS}=\frac1{8\pi}\oint{\tilde*}(\chi_1+\bar\chi_1-\rho_1-\rho'_1).
\end{equation}
Again one may check the Schwarzschild case: $\chi_1=-M$, $\rho_1=\rho'_1=-2M$,
yielding $E_{BS}=M$. Note that generally {\em the limit of $E_{BS}$ at spatial
infinity exists uniquely} and is given by the same formula (\ref{BS}). This
follows from the definition of functions regular at $\Im$ (\ref{regular}). Any
such transverse surface integral of functions regular at $\Im$ has a unique
limit at spatial infinity.

On the other hand, the ADM energy \cite{ADM} at spatial infinity can be
expressed as \cite{G,AH,AM,A,qle}
\begin{equation}
E_{ADM}=-\lim_{\omega\to0}\lim_{\omega'\to0}\frac{A^{1/2}}{(4\pi)^{3/2}}
\oint{*}\frac{\Re\Psi_2}{\chi\bar\chi}
\end{equation}
where, in the original treatment, it was unclear whether the limit depended on
the boost direction $\omega/\omega'$, i.e.\ on the choice of spatial
hypersurface. As above, this can be written
\begin{equation}
E_{ADM}=\lim_{\omega\to0}\lim_{\omega'\to0}\frac{A^{1/2}}{(4\pi)^{3/2}}
\oint{*}\frac{K+\rho\rho'-\Re(\sigma\sigma')}{\chi\bar\chi}
\end{equation}
which is similar to the expression (\ref{BS}) for the Bondi-Sachs energy, but
apparently differs by the term $\Re(\sigma\sigma')$ in the shears. The
discrepancy in the two energies is found from the asymptotic regularity
conditions (\ref{shear0}--\ref{shear1}) to be $O(\omega+\omega')$. This
discrepancy is generally non-zero at null infinity ($\omega=0$ or $\omega'=0$),
so that the ADM energy, if extended to null infinity by the same formula, would
generally not agree with the Bondi-Sachs energy. However, the discrepancy
vanishes at spatial infinity ($\omega=\omega'=0$) from any direction. Thus {\em
the ADM energy is the limit of the Bondi-Sachs energy at spatial infinity} in
this context, and also the limit of the Hawking energy from any spatial or null
direction:
\begin{equation}
E_{ADM}=\lim_{\omega\to0}\lim_{\omega'\to0}E=\lim_{\omega+\omega'\to0}E_{BS}.
\end{equation}
This provides a remarkably simple resolution of the long-standing questions
over the relation of the Bondi-Sachs and ADM energies, and the uniqueness of
the latter. In the current framework, one may simply use the Bondi-Sachs energy
on the entire asymptotic light cone $\Im$. The resolution explicitly rests on
the additional structure at spatial infinity provided by the advanced and
retarded conformal factors $(\omega,\omega')$.

Returning to the energy flux of gravitational radiation: the propagation
equations for the Hawking energy can be found from the compacted
spin-coefficient equations (4.12.32), using $\thorn{*}1=-{*}2\rho$ and
$\thorn'{*}1=-{*}2\rho'$, as \cite{mon}
\begin{eqnarray}
\thorn E&=&\frac{A^{1/2}}{(4\pi)^{3/2}}\left[\oint{*}\rho
\frac{K+\rho\rho'-\tau'\bar\tau'+\eth\tau'-\Phi_{11}-3\Pi}{\chi\bar\chi}
\right.\nonumber\\
&&{}+\left.\oint{*}\rho'\frac{\sigma\bar\sigma+\Phi_{00}}{\chi\bar\chi}
-\frac1A\oint{*}\rho\oint{*}\frac{K+\rho\rho'}{\chi\bar\chi}\right]\\
\thorn'E&=&\frac{A^{1/2}}{(4\pi)^{3/2}}\left[\oint{*}\rho'
\frac{K+\rho\rho'-\tau\bar\tau+\eth'\tau-\Phi_{11}-3\Pi}{\chi\bar\chi}
\right.\nonumber\\
&&{}+\left.\oint{*}\rho\frac{\sigma'\bar\sigma'+\Phi_{22}}{\chi\bar\chi}
-\frac1A\oint{*}\rho'\oint{*}\frac{K+\rho\rho'}{\chi\bar\chi}\right].
\end{eqnarray}
Rewriting these equations using the conformal spin-weighted derivatives
$(\hat\thorn,\hat\eth,\hat\eth',\hat\thorn')$, assuming either vacuum or
suitable fall-off of the matter fields, and using the asymptotic regularity
conditions and expansions (\ref{shear0}--\ref{exp1}), it can be shown that
$(\hat DE,\hat D'E)$ are $O(1)$, as expected on physical grounds. The general
argument is non-trivial, subtly confirming the asymptotic regularity
conditions, and uses the fact that $u$ is constant on the transverse surfaces
to cancel terms in $K+\rho\rho'$, and $\oint{*}\eth'\tau=0$ (4.14.69).
Fortunately the cases of most interest are $(\hat DE,\hat D'E)$ at
$(\Im^-,\Im^+$) respectively, where it is easier to see that most of the terms
disappear, in the vacuum case leaving just
\begin{eqnarray}\label{massgain}
\sqrt2\hat DE&=&\frac1{4\pi}\oint{\tilde*}|\sigma_1|^2 \quad\hbox{at $\Im^-$
($\omega'=0$)}\\\label{massloss} \sqrt2\hat
D'E&=&-\frac1{4\pi}\oint{\tilde*}|\sigma'_1|^2 \quad\hbox{at $\Im^+$
($\omega=0$)}
\end{eqnarray}
where
\begin{eqnarray}
\sigma_1&=&\sigma/u\omega\\
\sigma'_1&=&\sigma'/u\omega'
\end{eqnarray}
are the leading-order terms in the shears (\ref{shear0}--\ref{shear1}), regular
at $\Im$. In terms of the retarded and advanced news functions
\begin{eqnarray}
N&=&\sigma/u\\
N'&=&\sigma'/u
\end{eqnarray}
this implies
\begin{eqnarray}
\sqrt2DE_{BS}&=&\frac1{4\pi}\oint{\tilde*}|N|^2 \quad\hbox{at $\Im^-$}\\
\sqrt2D'E_{BS}&=&-\frac1{4\pi}\oint{\tilde*}|N'|^2 \quad\hbox{at $\Im^+$.}
\end{eqnarray}
The second equation is the usual Bondi-Sachs energy-loss equation, showing that
the outgoing gravitational radiation carries energy away from the system. The
first equation similarly shows that ingoing gravitational radiation supplies
energy to the system. The new energy propagation equations
(\ref{massgain}--\ref{massloss}) improve on these equations by applying also in
the limit at $i^0$. Thus the change in energy from $i^0$ to a section of
$\Im^\pm$ is finite, as mentioned above. Physically this means that {\em the
ingoing and outgoing gravitational radiation decays near spatial infinity such
that its total energy is finite}. This property, assumed in addition to the
Ashtekar-Hansen definition of asymptotic flatness \cite{AH}, is known to imply
the coincidence of Bondi-Sachs and ADM energies at spatial infinity \cite{AM},
but here the property has been derived from the new definition of asymptotic
flatness.

It is straightforward to generalize to an asymptotic energy-momentum vector
\begin{equation}
P=\lim_{u\to0}\frac{A^{1/2}}{(4\pi)^{3/2}}
\oint{*}\ell\frac{\sigma\sigma'-\Psi_2}{\chi\bar\chi}
\end{equation}
where the vector $\ell$ in Cartesian coordinates is (1.4.11)
\begin{equation}
\ell=(t,x,y,z)=\left(1,\frac{\zeta+\bar\zeta}{\zeta\bar\zeta+1},
\frac{i(\bar\zeta-\zeta)}{\zeta\bar\zeta+1},
\frac{\zeta\bar\zeta-1}{\zeta\bar\zeta+1}\right).
\end{equation}
Then $P$ coincides with the Bondi-Sachs energy-momentum \cite{BBM,S} at null
infinity and with the ADM energy-momentum \cite{ADM} at spatial infinity. The
corresponding energy-momentum propagation equations are
\begin{eqnarray}
\sqrt2\hat DP&=&\frac1{4\pi}\oint{\tilde*}\ell|\sigma_1|^2 \quad\hbox{at
$\Im^-$ ($\omega'=0$)}\\ \sqrt2\hat
D'P&=&-\frac1{4\pi}\oint{\tilde*}\ell|\sigma'_1|^2 \quad\hbox{at $\Im^+$
($\omega=0$)}
\end{eqnarray}
cf.\ \cite{NT}.

Finally, it should be noted that, while the gravitational radiation at
$(\Im^-,\Im^+)$ is respectively encoded as above in complex functions
$(\sigma_1,\sigma'_1)$ or $(N,N')$, a more tensorial formulation would encode
such shear terms in transverse traceless tensors, exactly as in the
conventional treatment of linearized gravitational radiation.

\section{Conclusion}
A new framework for space-time asymptotics has been proposed and developed,
refining Penrose's conformal framework by introducing advanced and retarded
conformal factors. This allows a relatively simple definition of asymptotic
flatness at both spatial and null infinity. It has been shown how to fix the
light cone at infinity so that it is locally a metric light cone. Assuming
smooth embedding of the light cone, asymptotic regularity conditions have been
proposed and asymptotic expansions developed. These conditions ensure that the
Bondi-Sachs energy-momentum is finite and tends uniquely to the uniquely
rendered ADM energy-momentum at spatial infinity, that the ingoing and outgoing
gravitational radiation has the expected two modes as encoded in gravitational
news functions, and that the energy-flux integrals decay at spatial infinity
such that the total energy of the gravitational radiation is finite. The most
basic physical properties of isolated gravitational systems have thereby been
included.

Mathematically the new structure proposed for infinity is unprecedentedly
rigid, allowing physical fields to have very simple behaviour at spatial
infinity. Currently there are no indications that this is too simple to be
physically realistic, though open issues remain. The asymptotic symmetry group
presumably simplifies accordingly.

The framework, implemented in the spin-coefficient or null-tetrad formalism, is
quite practical, allowing explicit calculations as for null infinity alone.
While this article has presented only a leading-order analysis, asymptotic
expansions can now be developed to higher orders. This would presumably allow
insights into angular momentum and multipole moments of the gravitational
field. A related issue is the asymptotic behaviour of the conformal curvature
near spatial infinity. It may even be possible to address the long-standing
question of finding a conformal form of the Einstein equations which is regular
at both null and spatial infinity. In any case, it is to be hoped that the
refined conformal picture will stimulate a renewal of interest in space-time
asymptotics.

\end{document}